\documentclass[aps,prl,showpacs,preprint,floatfix]{revtex4}
\usepackage{latexsym}
\usepackage{amsfonts}
\usepackage{amssymb}
\usepackage{amsmath}
\usepackage{graphicx}

\begin{document}

\title{Impact of the electrical connection of N Spin Transfer Oscillators on their synchronization : an analytical study}

\author{B. Georges, J. Grollier, V. Cros, A. Fert}
\affiliation{Unit\'e Mixte de Physique CNRS/Thales, Route
d\'epartementale 128, 91767 Palaiseau Cedex and Universit\'e
Paris-Sud XI, 91405, Orsay, France}

\date{\today}

\begin{abstract}
Spin Transfer Nano-Oscillators (STNOs) are good candidates for applications to telecommunications, but their output power is very low, under the nW. A possible solution to overcome this problem is to synchronize an assembly of STNOs. For manufacturing purposes, large number of STNOs will have to be electrically connected. In this letter, we study analytically the impact of electrical connection on the synchronization of STNOs. Our analysis shows that the phase dynamics of the coupled STNOs can be described in the framework of the Kuramoto model. The conditions for successful synchronization of an assembly of STNOs are derived.
\end{abstract}

\maketitle

The spin transfer torque \cite{Slonc,Berger} in nanometre-scale magnetic devices is a result of spin angular momentum  transfusion from a spin-polarized current to a ferromagnet. This effect can be used to induce microwave steady-state magnetization precessions in spin-valves or magnetic tunnel junction by injection of a dc current \cite{Tsoi,Kiselev2003,Rippard2004,KrivoBerkov,Nazarov,FuchsMgO}. Due to their tunability, high frequency emission, quality factor and sub-micron size, spin transfer nano-oscillators (STNOs) are good candidates for applications to telecommunications. The challenge is now to overcome the problem linked to their low emitted power, typically less than 1 nW. A solution is to achieve the synchronization of assemblies of STNOs, leading to coherent emission and increased power, as can be the case in arrays of Josephson junctions \cite{JainJosephson}. Mutual phase-locking between STNOs is possible due to their intrinsic non-linear behavior under the condition that their magnetization precessions are coupled. Several coupling mechanisms have been proposed.  Local coupling mediated by spin-waves, specific to the nanocontact geometry, has been demonstrated effective for synchronization of two STNOs \cite{Mancoff,Kaka}. It has been shown by zero kelvin macrospin simulation that the non-local coupling induced by the self-emitted microwave currents could also lead to mutual synchronization \cite{GrollierSync}. Although synchronization by this last mechanism has not yet been reported, phase-locking of a single STNO by an external microwave current has been evidenced \cite{RippardPRLinjection,BGeorges,Zhang}. In all cases, a large assembly of STNOs will require a common bias current. 

In this letter, we analyze the impact of electrical connection of N STNOs on their synchronization. In a previous paper, we have analytically studied the coupling efficiency of a single STNO to a microwave source \cite{BGeorges}. We extend this case to N electrically connected STNOs in interaction with their self-emitted microwave currents. We find that in the case of connections in series or in parallel, the final equations can be analysed with the Kuramoto model \cite{Kuramoto}. We then discuss the power emitted by any assembly of electrically connected STNOs. 
 
We first apply the theory of weakly forced oscillators to the case of STNOs. As in Ref.\cite{BGeorges}, we start from the equation for the amplitude of the spin wave mode derived by Slavin \textit{et al.} \cite{SlavinIEEE} including the spin transfer torque. From this equation of motion, it is possible to derive the expression of the phase $\Phi$ of the oscillator. The perturbations to the limit cycle are assumed to be small. Following the procedure of Ref.\cite{Pikovski}, we found in Ref.\cite{BGeorges} the Adler equation \cite{Adler} for the phase dynamics of a STNO forced by an external microwave current. From this derivation an expression for the normalized coupling strength between the oscillator and the source, $\epsilon$/I$_{hf}$, can be obtained as a function of experimentally available parameters : I$_{dc}$, I$_{th}$ and the agility in current $\frac{\partial f_0}{\partial I_{dc}}$:
\begin{equation}\label{epsilon}
\frac{\epsilon}{I_{hf}} = \frac{\sigma tan(\gamma)}{2\sqrt{2}} \sqrt{\frac{I_{dc}}{I_{dc} - I_{th}}} \sqrt{1 + \left(\frac{2 \pi I_{dc}}{\sigma I_{th}} \frac{\partial f_0}{\partial I_{dc}}\right)^2}
\end{equation}
Experiments of phase locking between an STNO and a microwave source showed very good agreement with this model and enabled an experimental determination of $\epsilon$/I$_{hf}$ \cite{BGeorges}.

\begin{figure}[h]
   \centering
    \includegraphics[keepaspectratio=1,width=8.5 cm]{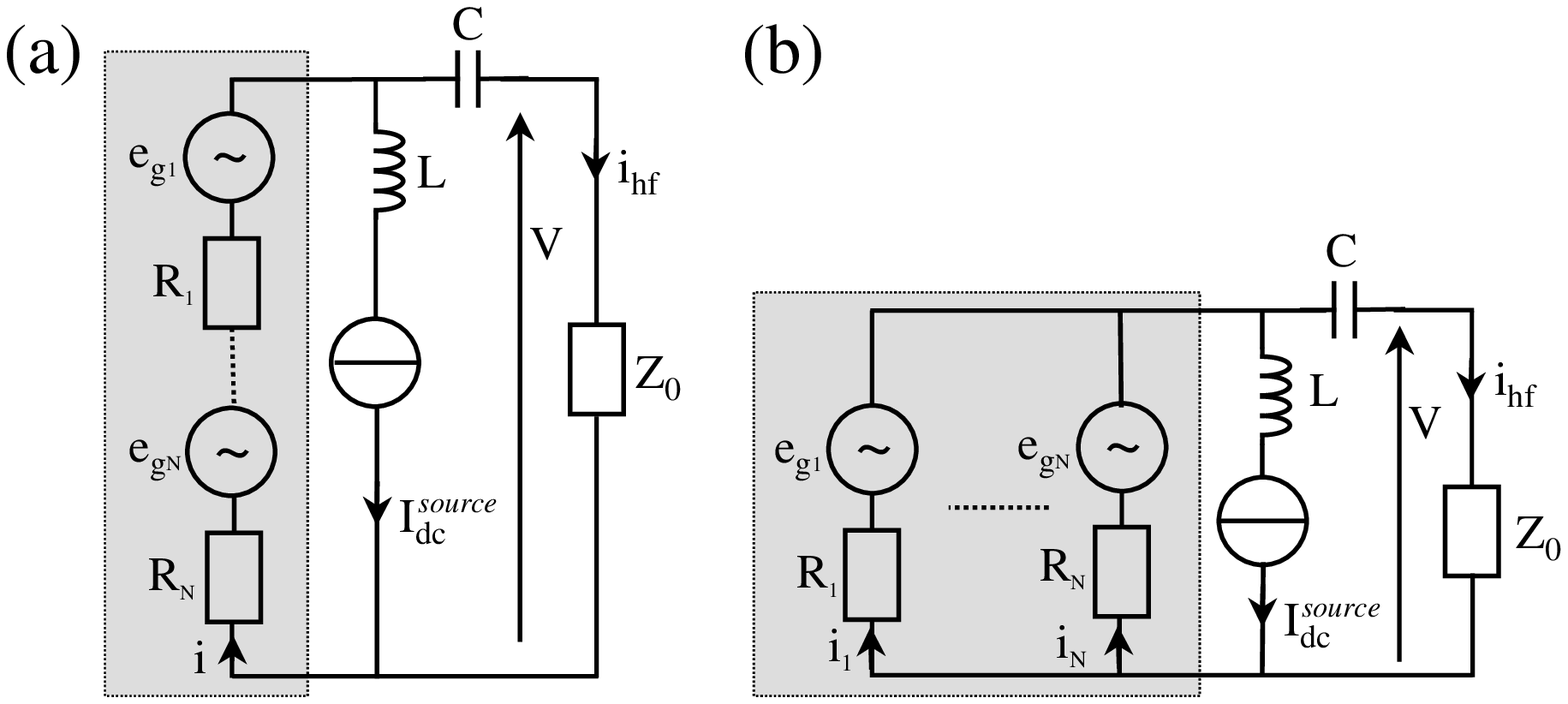}
     \caption{Georges et al.}
    \label{fig1}
\end{figure}

In the following of the letter, we extend our derivation of the phase dynamics for a single STNO coupled to an external microwave signal to the case of an STNO assembly coupled by their self-emitted microwave currents.
For the sake of simpleness, we consider that all STNOs have the same resistance R, and that precessions lead to the same resistance variation $\Delta R_{osc}$. Each STNO n produces a microwave voltage e$_{g}(n)$ = $\Delta R_{osc}$I$_{dc}$$cos(\varphi_{n})$. We first concentrate on the case of a series connection of N STNOs to a load Z$_{0}$, as illustrated in Fig.\ref{fig1}(a). The inductance and capacitance in the circuit decouple the microwave and dc currents. The microwave current i$_{hf}$ circulating in the loop is :
\begin{equation}\label{ihfseries}
\left(i_{hf}\right)_{series} = -\frac{\Delta R I_{dc}}{Z_0 + NR} \sum^{N}_{n=1}  cos(\varphi_n)
\end{equation} 
By following the same procedure that enabled us to derive Adler's equation in Ref.\cite{BGeorges}, we obtain the equation for the phase of the oscillator number n :
\begin{equation}\label{kuramoto}
\frac{d(\Phi_n)}{dt} = - 2 \pi f^{n}_{0} - \frac{K}{N} \sum^{N}_{i=1}  cos(\Phi_i - \Phi_n+ \Phi_0) + \xi_n(t)
\end{equation} 
This is the Kuramoto equation \cite{Kuramoto}, with the coupling factor K proportional to the current i$_{hf}$ in the loop and the coupling efficiency $\epsilon$/$I_{hf}$.
\begin{equation}\label{Kseries}
K_{series} = \left(\frac{\epsilon}{I_{hf}}\right) \frac{N}{Z_0+NR}\Delta R I_{dc}
\end{equation} 
Kuramoto has shown that Eq.\ref{kuramoto} can be solved analytically, assuming that the number of oscillators is large and that the frequency distribution is lorentzian with a width at half maximum D$^2$. The noisy terms are supposed to be gaussian with zero mean, $\delta$ correlated in time and independent for each oscillator : $\left\langle \xi_n\right\rangle = 0, \left\langle \xi_m(t)\xi_n(t')\right\rangle = 2 w^2 \delta(t-t') \delta_{mn}$. According to the Kuramoto model, synchronization is possible if the order parameter K becomes larger than the critical coupling K$_c$ = 2(w$^2$+D$^2$). 

Eq.\ref{Kseries} is very similar to the one obtained for N Josephson junctions connected in series \cite{StrogatzJosephson}. The main difference is that the resistance of superconducting Josephson junctions is extremely small : the coupling parameter K increases with the number of junctions. An STNO resistance is typically 10 $\Omega$ for all-metallic structures, and 200 $\Omega$ for MgO tunnel junctions. Already for a small number of oscillators, the resistance NR becomes larger than the typical value $Z_0$ = 50 $\Omega$. In the case of STNOs, for large N, according to Eq.\ref{Kseries}, the coupling does not increase with N. The magnetoresistance of each oscillator will need to be large enough to synchronize them all. Writing the condition  K $>$ K$_c$ provide two requirements for the transition to synchronization:
\begin{eqnarray}
\left(\frac{\Delta R_{osc}}{R}\right)_{series}&>&\left(\frac{\Delta R_{osc}}{R}\right)_{th} = \frac{2(\gamma ^2+w^2)}{I_{dc}}\frac{1}{\left(\frac{\epsilon}{I_{hf}}\right)} \label{CondSyncSeries} \\ 
N_{series}&>&\frac{\left(\frac{\Delta R_{osc}}{R}\right)_{th} \frac{Z_0}{R}}{\frac{\Delta R_{osc}}{R} - \left(\frac{\Delta R_{osc}}{R}\right)_{th}} \label{CondSyncSeries2}
\end{eqnarray} 
Eq.\ref{CondSyncSeries} gives a condition on the ratio $\Delta R_{osc}$/R. For state-of-the-art STNOs, it is now possible to reach a frequency dispersion D$^2$ of 100 MHz and a linewidth w$^2$ of 10 MHz, with a biasing current I$_{dc}$ of the order of 5 mA. From our experiments on injection locking with a microwave source, we have been able to determine that for an agility of 70 MHz/mA, the coupling efficiency $\epsilon$/I$_{hf}$ is about 35 MHz/mA \cite{BGeorges}. Experiments and calculations (Eq.\ref{epsilon}) show that the coupling efficiency increases with the agility in current. As STNOs can easily reach agilities up to 1GHz/mA \cite{RippardPRB2006}, the corresponding $\epsilon$/I$_{hf}$ ratio obtained from Eq.\ref{epsilon} is 300 MHz/mA.
From Eq.\ref{CondSyncSeries} the threshold $\Delta R_{osc}$/R ratio for synchronization is of the order of 15$\%$. Note that $\Delta R_{osc}$/R does not correspond to the total MR ratio, but to the part converted in precessions. For example MgO based tunnel nanopillars can reach MR ratios of 100 $\%$ but up to now the best emission peaks give a power of about 50 nW, which corresponds, taking R = 200 $\Omega$, to $\Delta R_{osc}$/R = 1e-5 \cite{Nazarov}. All metallic nanopillars are also far from fulfilling the conditions for synchronization since up to now their total magnetoresistance is lower than 10 $\%$. In a previous paper, we had predicted by macrospin numerical simulations that synchronization could occur for MR ratios lower than 5 $\%$ \cite{GrollierSync}. The discrepancy between our present result and the former lies in the fact that macrospin simulations predict much larger agilities in current (about 10 GHz/mA) than obtained experimentally \cite{Kiselev2003}. Once the first condition expressed in Eq.\ref{CondSyncSeries} is satisfied, the second requirement for synchronization, expressed in Eq.\ref{CondSyncSeries2} is easily fulfilled. For example by taking $\Delta R_{osc}$/R = 1.1 $\left(\Delta R_{osc}/R\right)_{th}$ and Z$_0$ = 10R, 100 oscillators are required for synchronization, which is commonly manufacturable. 

Possible routes to achieve synchronization by the self-emitted microwave currents coupling are i) reducing the frequency dispersion of the assembly ii) increasing STNOs agility iii) achieving large angle magnetization precessions in MgO based tunnel junctions. Following route i) let us consider a frequency dispersion of 10 MHz with the other parameters kept fixed. Then the threshold for synchronization is reduced to $\Delta R_{osc}$/R = 2.6 $\%$, that can be reached even in metallic structures. Increasing agility as suggested in ii) can be obtained by achieving large angle excited modes close to the uniform mode. This goes with structures of reduced diameter, high current polarization, and biasing with an angle of the reference layer. 

The two conditions in Eq.\ref{CondSyncSeries} and Eq.\ref{CondSyncSeries2} only give the threshold for synchronization. Coherent emission of the entire STNO array requires higher coupling values \cite{Pikovski}. Moreover, delays in the transmission lines can hinder the coupling \cite{Akerman}. Analytical solutions to the extended Kuramoto equation with delay have been derived \cite{StrogatzDelay}. The best conditions for synchronization correspond to values of the delay multiples of the precession period.

We now discuss the emitted power that can be achieved if all the oscillators are phase-locked. For the series and parallel connections with  $I_{dc}$ the dc current in each branch, if all N oscillators are synchronized, the microwave power delivered to the load Z$_0$ is : 
\begin{equation}\label{powerseries}
P_{series,parallel} = \frac{Z_0 N^2}{Z_{series,parallel}^2} \Delta R^2 I_{dc}^2 
\end{equation} 
with $Z_{series}=Z_0+NR$ and $Z_{parallel}=NZ_0+R$. In series, this means that if NR $>>$ Z$_0$ which is the case if the load Z$_0$ is fixed to the standard 50 $\Omega$, the power does not increase with the number of oscillators for large values of N. The case, illustrated on Fig.\ref{fig1}(b), of oscillators connected in parallel is similar. The STNOs connected in parallel tend to shunt each other, and the power will increase as N$^2$ only if NZ$_0$ $<<$ R. As for the series case, a compromise has to be done since a small Z$_0$ decreases the output power. The conditions for synchronization in the case of parallel connections are very similar to the series case. To obtain them, $Z_0/R$ in Eq.\ref{CondSyncSeries2} should be replaced by $R/Z_0$.

\begin{figure}[h]
   \centering
    \includegraphics[keepaspectratio=1,width=8.5 cm]{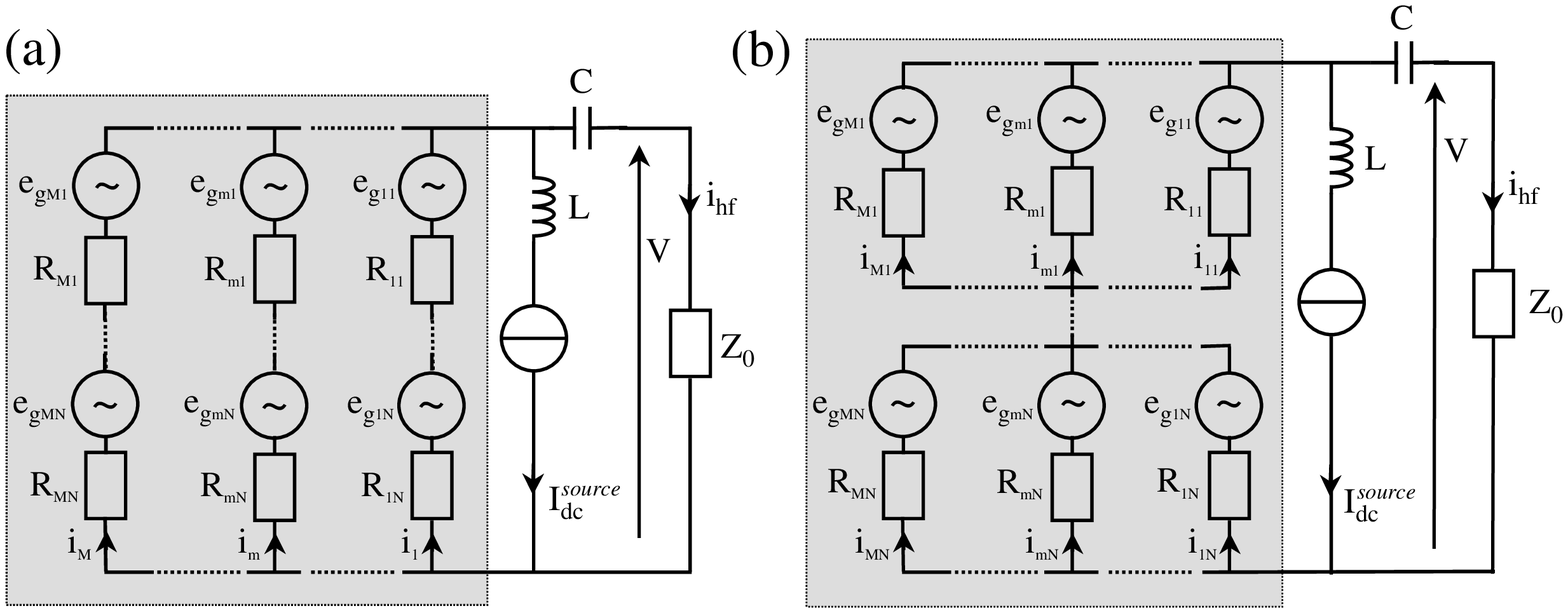}
     \caption{Georges et al.}
    \label{fig2}
\end{figure}

When the value of the load can be chosen, which is the case for on-chip applications, a solution for the use of series or parallel networks is to tune Z$_0$ (increase of Z$_0$ in series, decrease in parallel). By considering in series that Z$_0$ = 10 NR and in parallel that R = 10 Z$_0$N, then if all STNOs are synchronized :
\begin{equation}\label{powerZ0}
P_{series} = P_{parallel} \approx \frac{N}{10R} \Delta R^2 I_{dc}^2 
\end{equation} 
As can be seen from Eq.\ref{powerZ0}, the power increases as N if the entire assembly is synchronized. A good way to avoid matching impedance problems is to use hybrid arrays, such as represented on Fig.\ref{fig2}. Fig.\ref{fig2}(a) illustrates the case of M branches in parallel of N STNOs connected in series, and Fig.\ref{fig2}(b) the connection in series of N groups of M STNOs connected in parallel. In these cases, the total number of STNOs is NM. If all oscillators are synchronized, the output power of such arrays is (with $I_{dc}$ in each branch):
\begin{equation}\label{powerhybrid}
P_{hybrid} = \frac{N^2M^2 Z_0}{(NR+MZ_0)^2} \Delta R^2 I_{dc}^2 
\end{equation}
Then by choosing NR=M$Z_0$ the conditions of impedance matching are fulfilled and the power increases as the number of oscillators NM. In the case of the hybrid network of Fig.\ref{fig2}, following the procedure used for deriving Eq.\ref{kuramoto} we find that the phase of the oscillator (n,m) ($n^{th}$ STNO of branch number m) is ruled by the following equation:
\begin{eqnarray}\label{kuramotohybrid}
\frac{d(\Phi_{n,m})}{dt} = - 2 \pi f^{n,m}_{0} + \frac{K_1}{N} \sum^{N}_{i=1}  cos(\Phi_{i,m}-\Phi_{n,m}+\Phi_0) \\ \nonumber
- \frac{K_2}{NM} \sum^{N}_{i=1} \sum^{M}_{j=1} cos(\Phi_{i,j}-\Phi_{n,m}+\Phi_0) + \xi_{n,m}(t)
\end{eqnarray} 
where:
\begin{equation}\label{Khybrid}
K_1 =  \left(\frac{\epsilon}{I_{hf}}\right) \frac{\Delta R_{osc}}{R} I_{dc} , K_2  = \left(\frac{\epsilon}{I_{hf}}\right) \frac{Z_0}{R} \frac{M \Delta R_{osc}}{MZ_0+NR} I_{dc}
\end{equation} 
Eq.\ref{kuramotohybrid} corresponds to an extended version of the Kuramoto equation, to which we do not know any simple analytical solution. For large values of N (and M), K$_1$ and K$_2$ are again independent on the number of oscillators. Due to the similarity with Eq.\ref{kuramoto}, we believe that the conditions for synchronization will be very close to those given in Eq.\ref{CondSyncSeries} and \ref{CondSyncSeries2}.

The preceding considerations for the emitted output power of the array are valid for all type of couplings, because STNOs assemblies will always be electrically connected. The best solutions concerning the power issue are combinations of series and parallel electrical connection of STNOs. We point out that the nanocontact geometry that goes with the local spin-wave coupling, is hardly suitable for series connection. 

In summary, we have analytically studied the synchronization effect by the self-emitted microwave currents in electrically connected arrays of spin transfer nano-oscillators. In this scheme, STNOs are well described by the Kuramoto model, from which the conditions for succesful synchronization are derived. Having extracted from our experiments the value of the coupling to a microwave current, we give the requirements on the microwave characteristics of STNOs for phase locking. The output power of the arrays has also been calculated in the case of complete synchronization. From this last consideration, that can be extended to all type of couplings, the best connections schemes have been identified : hybrid arrays. 

This work was partly supported by the French National Agency of Research ANR through the PNANO program (NANOMASER PNANO-06-067-04) and the EU network SPINSWITCH (MRTN-CT-2006-035327).

\newpage

\textbf{Figure captions}\\

Figure 1. Scheme for STNO connections to the load Z$_{0}$ (a) in series (b) in parallel \\

Figure 2. Scheme for STNO connections to the load Z$_{0}$ (a) hybrid 1 (b) hybrid 2 \\

\newpage

\newpage

\end{document}